\documentclass[prl,twocolumn,showpacs,preprintnumbers,amsmath,amssymb]{revtex4}
\usepackage{graphicx,lscape}
\usepackage{dcolumn}
\usepackage{bm,txfonts}
\usepackage{color}

\begin{document}
\title{Emission properties of an oscillating point dipole from a gold Yagi-Uda nanoantenna array}

\author{S.~V.~Lobanov$^{1,2}$, T.~Weiss$^3$,  D.~Dregely$^4$,  H.~Giessen$^4$, N.~A.~Gippius$^{2,5}$, and S.~G.~Tikhodeev$^{1,2}$}

\affiliation{
  $^1$ M. V. Lomonosov Moscow State University, Leninskie Gory 1, Moscow 119991, Russia\\
  $^2$ A. M. Prokhorov General Physics Institute, Russian Academy of Sciences, Vavilova Street 38, Moscow 119991, Russia\\
  $^3$ Max Planck Institute for the Science of Light, G\"{u}nter-Scharowsky-Stra{\ss}e 1\/Bau 24, D-91058 Erlangen, Germany\\
  $^4$ 4$^\mathrm{th}$ Physics Institute and Research Centers Scope and Simtech, University of Stuttgart,
  D-70550 Stuttgart, Germany\\
  $^5$ LASMEA, University Blaise Pascal, 24 Avenue des Landais, F-63177 Aubi\`{e}re Cedex, France
}

\begin{abstract}
We investigate numerically the interaction of an oscillating point dipole with a periodic array of optical
Yagi-Uda nanoantennas in the weak coupling limit.
A very strong near-field enhancement
of the dipole emission by the resonant plasmon
mode in the feed element is predicted in this structure. It is shown that the
enhancement strength depends strongly on the dipole position, the direction of the dipole
moment, and the oscillation frequency. The radiative intensity of the point dipole
from appropriate places next to one feed element may exceed the radiative intensity of an
equivalent dipole in free-space by a factor of hundred. In spite of only one
director used in each  nanoantenna of the array, the far-field emission pattern is highly
directed. The radiative efficiency (the ratio of the radiative to the full emission) appears  to be around 20\%.
\end{abstract}
\date{30 September 2011}

\maketitle

\renewcommand{\baselinestretch}{1.2}

Nanophotonics has been in the focus of intensive investigations in recent years. One of its numerous
areas is the nanoscale controlling of light emission from a single molecule or quantum
dot. Promising tools for the realization of this goal are optical
nanoantennas~\cite{Gersen2000,Li2007,Hofmann2007,Kuhn2008,Taminiau2008,Mohammadi2008,Li2009,Sun2008,Sun2009,Kinkhabwala2009,Esteban2010,Koenderink2009,Kosako2010}.

 Metal antennas are traditionally used for controlling the radiation pattern of
electromagnetic wave emission in the radio and microwave frequency range.
Though the electromagnetic properties of metals in the optical range differ significantly from that  in the
radio and microwave range, it seems to be reasonable to use the main concepts of radio
antennas in the optical range as well. It has been suggested~\cite{Li2007,Hofmann2007,Taminiau2008,Li2009,Kosako2010} to
construct a nanooptical antenna with elements that are arranged as in the radio Yagi-Uda antennas.

Yagi-Uda antenna consists usually of one or two reflectors, one feed element and
several directors with appropriately selected scattering phases (reflector and  director are slightly detuned inductively and
capacitively). As it has been recently shown, the nanoantenna elements can be
nanorods~\cite{Hofmann2007,Taminiau2008,Kosako2010}, core-shell~\cite{Li2007,Li2009}
or spherical~\cite{Koenderink2009}  nanoparticles. All
elements scatter the light and the resulting interference forms a highly directed beam along the antenna axis.
The size of the elements in such optical Yagi-Uda nanoantennas have to be smaller than the light emission
wavelength in free-space. Such optical nanoantennas only work efficiently in narrow frequency domains,
where the interaction of the emitter with light is resonantly enhanced because of the excitation of so-called localized
plasmons~\cite{Barnes2003,Polman2008} in the nanoantenna elements.

Spontaneous emission is not an intrinsic atomic property, but it depends sensitively on
the local density of photonic modes at certain
frequencies in a microcavity~\cite{Purcell1946,Ford1984}, or, equivalently, on the local electromagnetic field
value at the position of the quantum emitter~\cite{Kuhn2008,Mohammadi2008,Esteban2010}.
Using resonances, it is possible to increase the local electromagnetic field significantly, and, resultantly, to enhance and redirect
the dipole emission.
In the case of  localized plasmon resonances, the collective excitation of electrons at the plasmon
frequency leads to a
considerably enhanced emission rate when the point dipole is located in the
vicinity of metallic nanoparticles with the appropriate orientation of its dipole moment~\cite{Taminiau2008,Sun2008,Kinkhabwala2009}.

The exact description of photon radiation from the quantum emitter located in some
metal-dielectric environment is very complicated. A convenient approximation is a model of an
oscillating point dipole. It oscillates with constant frequency and magnitude fixed by the
external source (so-called weak coupling limit). In other words, the emission
of a current $\vec{j}(\vec{r},t)=\vec{j}_0 \cdot \delta(\vec{r}-\vec{r}_0) \cdot e^{-i\omega t}$
inside an environment with spatially modulated permittivity has to be calculated. This system is now classical
and can be described by Maxwell's equations.

The goal of this paper is to calculate the  radiation pattern and emission rate of one oscillating
point dipole located in the periodic array of optical Yagi-Uda nanoantennas. Each antenna in the array consists
of three rectangular gold elements, director, feed, and  reflector.
The point emitter is located a few nanometers from the edge of one  feed element.
To the best of our knowledge, the emission properties of an
oscillating point dipole from such a structure have not been investigated yet.
The so far investigated structures have been either single Yagi-Uda nanoantennas~\cite{Li2007,Hofmann2007,Taminiau2008,Li2009,Kosako2010},
or arrays of simpler spherical shapes~\cite{Sun2008}. Using  arrays of
antennas in combination with several emitters with controlled  phase difference of oscillation, we can
control the emission directivity additionally~\cite{Dregely}, as in phased antenna arrays.

The calculation of the radiation characteristics from oscillating dipoles in two-dimensional
photonic crystal slabs can be performed using the method
of scattering matrix~\cite{Whittaker1999,Taniyama2008}.
Its main idea is to divide the structure into two parts adjacent to a plane through the dipole,
to find the scattering matrices of these parts, and to combine them by considering the boundary
conditions at the plane with the dipole. We use the Fourier modal method
(FMM)~\cite{Whittaker1999,Tikhodeev2002}, which can be improved by the
formulation of the correct Fourier factorization rules~\cite{Li2003} and adaptive spatial
resolution~\cite{Granet2002}, as well as of matched coordinates for more complex
metallic shapes~\cite{Weiss2009}.


\begin{figure}
\includegraphics[width=0.9\linewidth]{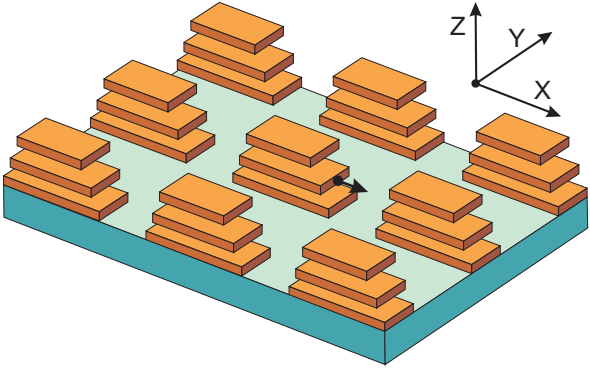}
\caption{(Color online) Lateral view of a periodic array of optical Yagi-Uda nanoantennas with an $x$-oriented
oscillating point dipole (black point with arrow) located at a horizontal distance of 5~nm from the feed element of one nanoantenna.} \label{Structure}
\end{figure}

The structure of interest is shown schematically in Fig.~\ref{Structure}. It consists of a
glass superstrate ($\varepsilon=2.4$), a periodic array of optical Yagi-Uda
nanoantennas ordered in a rectangular lattice, and a quartz substrate
($\varepsilon=2.13$). The periods along the $x$- and $y$-axis are equal to 450~nm and 300~nm, respectively.
Each antenna consists of three rectangular gold parallelepipeds
of 30~nm height and 100~nm width. The lengths of the top (director),
middle (feed), and bottom (reflector) elements are 220, 250 and 300~nm,
respectively. They are located in glass and the vertical distance between them is equal
to 100~nm. We assumed the gold permittivity to be described by the Drude formula with
9016~meV plasma frequency and 81~meV damping rate.

\begin{figure}[b]
\includegraphics[width=\linewidth]{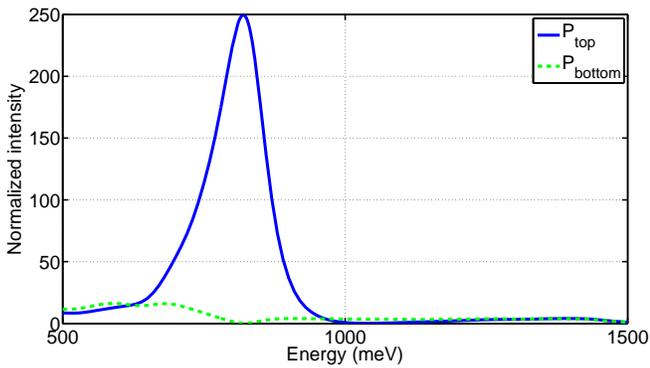}
\caption{(Color online) Calculated spectra of the emission  in top (blue curve) and bottom
(green curve) directions of  the oscillating point dipole (directed along $x$) coupled to the gold Yagi-Uda nanoantenna array.
The intensity is normalized to the maximum radiation intensity of an equivalent dipole (with the same
magnitude and the same frequency) in free space.}
\label{Spectra}
\end{figure}

Our first goal  is to calculate
the directional pattern as well as the emission spectra in the direction normal to the antenna plane. It
makes sense to normalize  the computed emission intensity $P(\vartheta,\varphi)$,
i. e.,  the Poynting vector of the dipole emission in the far field as a function of the spherical
angles $\vartheta$ and $\phi$, to the maximum intensity of the emission of a point dipole in free-space,
which oscillates with the same magnitude and frequency.
Thus, we can easily distinguish the enhancement of emission ($P>1$) from the attenuation ($P<1$) compared
to the dipole located in homogeneous vacuum.

\begin{figure*}
\begin{minipage}{.3\textwidth}
\includegraphics[width=\linewidth]{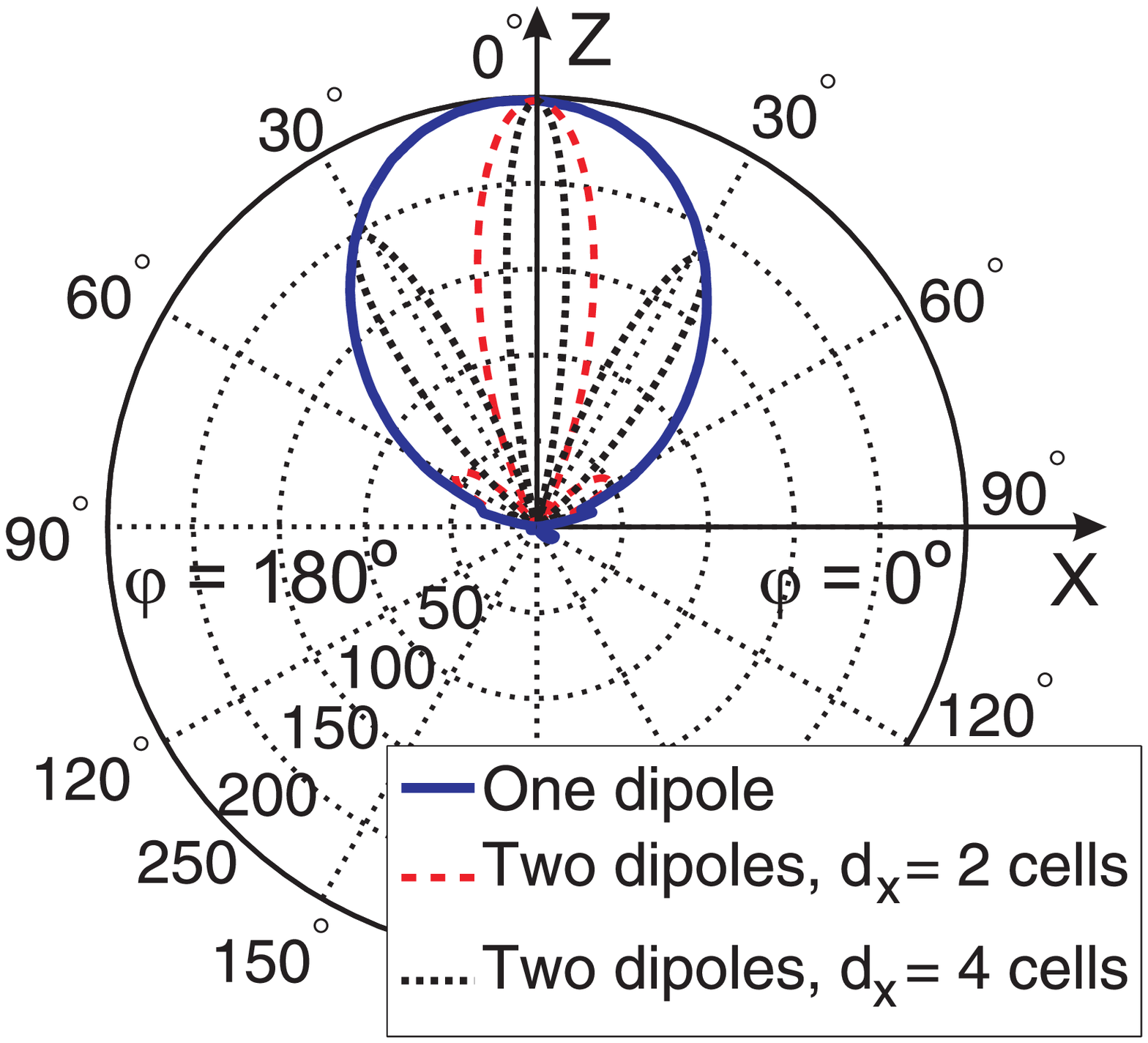}
\end{minipage}
\begin{minipage}{.32\textwidth}
\includegraphics[width=\linewidth]{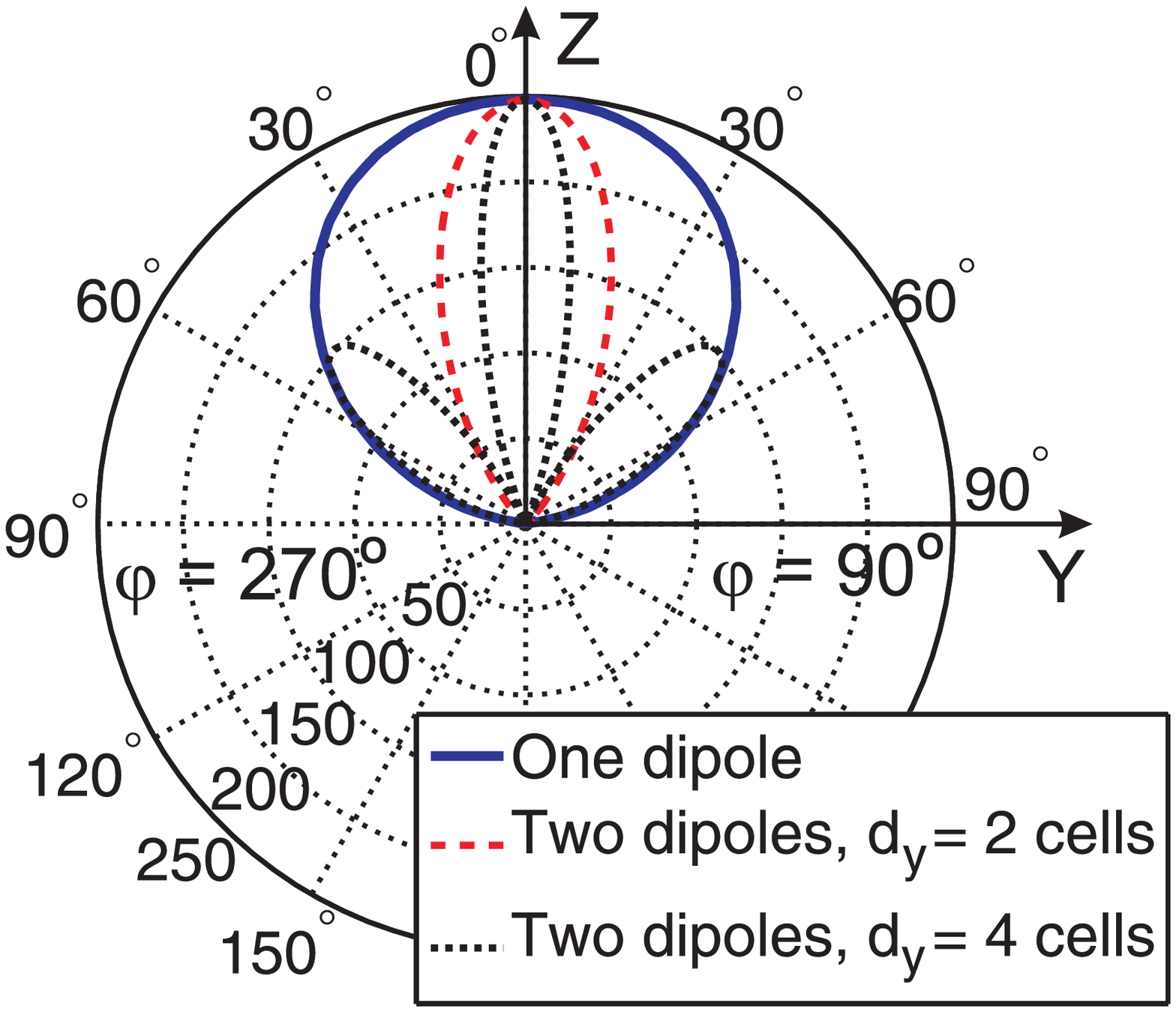}
\end{minipage}
\begin{minipage}{.3\textwidth}
\includegraphics[width=\linewidth]{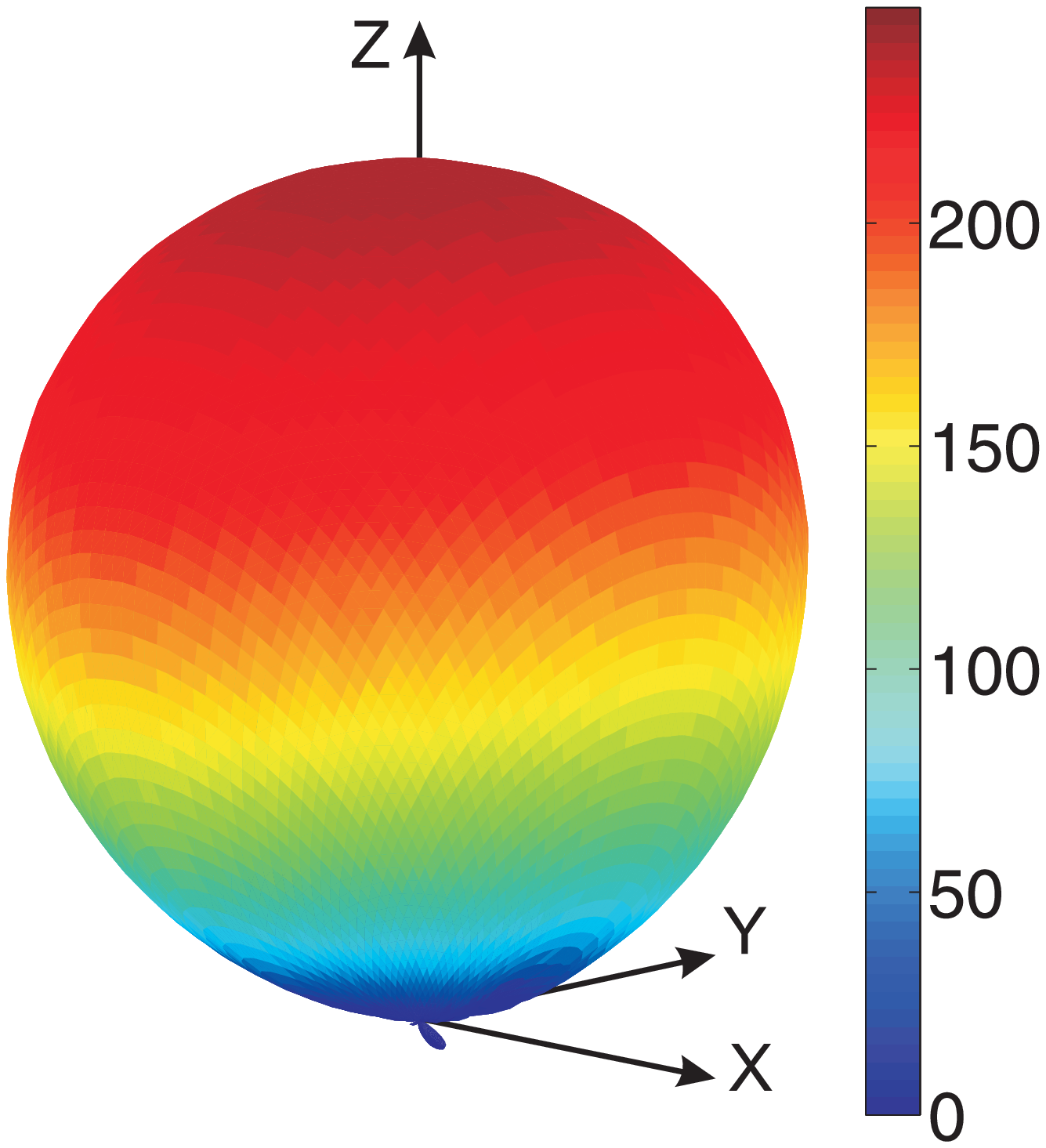}
\end{minipage}
\caption{(Color online) Calculated  2D polar diagrams (solid lines) of  far-field emission in $xz$ (left panel) and $yz$ (central panel) planes, as well as
the full 3D-directional diagram (right panel)
for the $x$-oriented oscillating point dipole coupled to the gold Yagi-Uda nanoantenna array at
resonant photon energy ($\hbar \omega = 820$~meV). The emission intensity is normalized
to the maximum radiation intensity of the equivalent dipole (with the same magnitude and the same frequency) in free space
and shown in the right panel as colored surface
$P(\vartheta,\varphi)\,\vec{\mathrm{e}}_r$. The dipole position is in the center-of-coordinates.
Dashed lines in left and central panels show the scaled by a factor of $1/2$ polar diagrams of $ P_2(\vec{d},\vartheta, \varphi)$,
the emission of a pair of synchronized dipoles separated by distance $\vec{d}$ (see explanation in the text).
}
\label{Dir_Pat}
\end{figure*}

To characterize quantitatively the full dipole emission and its radiative part, it is convenient to define
the Purcell factor and its radiative part as
\begin{equation}\label{eq:FP}
F_\mathrm{P}=\frac{\oiint\limits_{\Sigma_0} (\vec{P}\cdot d\vec{A})}{\oiint\limits_{\Sigma_0} (\vec{P}_0\cdot d\vec{A})} \, , \,\,
\mathrm{and} \,\,
F_\mathrm{P}^\mathrm{rad}=\frac{\oiint\limits_{\Sigma} (\vec{P}\cdot d\vec{A})}{\oiint\limits_{\Sigma_0} (\vec{P}_0\cdot d\vec{A})} \, ,
\end{equation}
respectively. Here, $\Sigma_0$ and $\Sigma$ denote spheres  with an infinitely small and large radius,
respectively, that surround the point dipole. $\vec{P}$ is the Poynting vector of the dipole emission
from the antenna array, $\vec{P}_0$ is its counterpart for dipole emission in free space.
The antenna absorption losses can be characterized by the
non-radiative part of the Purcell factor $F_\mathrm{P}^\mathrm{nr}$,
\begin{equation}
F_\mathrm{P}^\mathrm{nr} = F_\mathrm{P}-F_\mathrm{P}^\mathrm{rad}\, .
\end{equation}
Now, we can also introduce the radiative efficiency $\eta$ indicating the radiative part of dipole emission,
\begin{equation}
\eta=\frac{F_\mathrm{P}^\mathrm{rad}}{F_\mathrm{P}}.
\end{equation}
In what follows the results of these quantities will be presented employing 1633 spatial harmonics in the FMM~\cite{Tikhodeev2002,Weiss2009} .

The calculated emission spectra
in top ($+z$) and bottom ($-z$) directions
of the $x$-directed
oscillating point dipole located inside the periodic array of gold Yagi-Uda
nanoantennas are shown in Fig.~\ref{Spectra}. The dipole is placed on the horizontal symmetry axis
along $x$-direction at a distance of 5 nm to the edge of the feed element (see in Fig.~\ref{Structure}).  Only one strong and narrow
 resonance occurs in the emission spectra in the top direction
 at the photon energy $\hbar \omega = 820$~meV  ($\lambda =1.5~\mu$m). Its
magnitude is about 250 and the FWHM is 107~meV. The emission to
the bottom direction is
significantly smaller.  It has even a minimum at the plasmon resonance.

Figure~\ref{Dir_Pat} depicts the calculated radiation pattern of emission $P(\vartheta,\varphi)\,\vec{\mathrm{e}}_r $  at the resonant photon energy
$\hbar \omega = 820$~meV (where $\vec{\mathrm{e}}_r $ is the radial unit vector of a spherical coordinate system centered in the
point dipole position).
The calculated  radiative part of the Purcell factor
according to the second Eq.~(\ref{eq:FP})
appears to be as high as
 80 (see also Fig.~\ref{Pf_rad} below), which indicates
a very strong enhancement of  the dipole emission by the feed element.

In spite of only
one director employed in our Yagi-Uda antenna, the emission in top direction appears to be
highly directional. In order to characterize it,
the angular directivity $D(\vartheta,\varphi)$ can be calculated~\cite{Balanis2005},
which indicates the part of the full emission radiated along the direction
$(\vartheta$, $\varphi)$,
\begin{equation}
D(\vartheta,\varphi)=\frac{4\pi P(\vartheta,\varphi)}{\iint P(\vartheta,\varphi)d\Omega}.
\end{equation}
Its maximum value $D^\mathrm{max} = \max[D(\vartheta,\varphi)]$ is called directivity and indicates
the antenna's ability to form a narrow beam. The larger the directivity, the narrower
is the light beam. An isotropic radiator would have a directivity of 1, whereas for an
oscillating point dipole in free space, $D^\mathrm{max}=1.5$. In the case of the radiation pattern of Fig.~\ref{Dir_Pat},
$D^\mathrm{max} = 4.7$, i.e., it exceeds the directivity of the dipole in free space by a factor of 3.1.
In~\cite{Taminiau2008} the directivity of a single Yagi-Uda antenna with
three directors has been calculated as 6.4, which is only 1.37 times larger than in our case with only a single director.
Furthermore, we can also increase the directivity using more than one dipole emitter coupled to different antenna array
elements which is shown below.

It is also instructive to investigate the dependence of the emission enhancement
on the dipole position and on the orientation of its dipole moment respectively to the feed element.
Figure~\ref{Pf_rad} shows the calculated dependence of the radiative part of the Purcell factor
$F_\mathrm{P}^\mathrm{rad}$ with respect to the distance between the feed element and the dipole
 for $x$-, $y$-, and
$z$-directed dipoles. The emission enhancement of the antenna array decreases with increasing distance.
The strong polarization and
distance dependence demonstrates the local nature of the
antenna-dipole enhancement.

\begin{figure}[t]
\includegraphics[width=\linewidth]{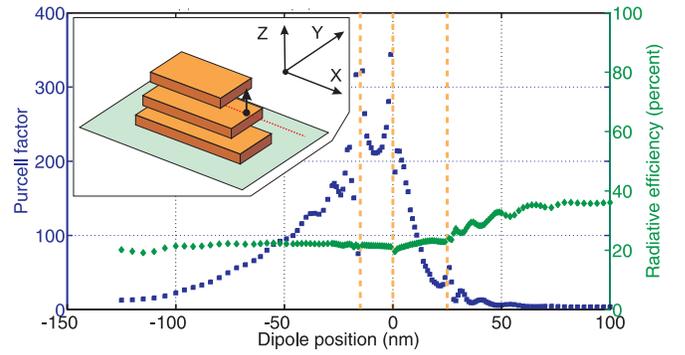}
 \caption{(Color online) Calculated dependencies of the Purcell factor $F_\mathrm{P}$ (squares)
 and the radiative efficiency $\eta$ (diamonds) on the dipole position for a $z$-polarized dipole.
 The $x$-coordinate of the dipole is changed from $ -125$~nm to 100~nm (measured from the position of the edge of the feed element),
 while the $y$- and $z$-coordinate are fixed.  The vertical dashed lines mark
the positions exactly of the edges of the director, feed, and reflector elements.
The geometry is explained in the
inset: the $z$-polarized dipole is shown as a black dot with arrow; it
lies 10~nm above the feed's upper horizontal surface,
is
centered with respect to the feed element along $y$-direction,
and shifted in $x$-direction, along
the dotted horizontal line.
}\label{Pf}
\end{figure}

For the different distances indicated in the inset of Fig.~\ref{Pf_rad}, the emission is only significantly enhanced for the $x$-polarized dipole,
which means that a plasmon mode with charges oscillating along $x$
is excited in the system. Such mode leads to a  very strong enhancement
of emission and determines the directional pattern. As the electromagnetic field near the
edges of the feed element is known to have a dipolar character at the fundamental plasmon mode, it is not surprising that
the radiation pattern of the $z$-polarized dipole displaced vertically by 10 nm above the
edge of the feed element (see the inset in Fig.~\ref{Pf}) nearly coincides with that of the $x$-polarized dipole 5 nm
apart from the edge of the feed element (see Fig.~\ref{Dir_Pat}). Unlike a $z$-polarized dipole in
free space, the majority of the emission is directed along the dipole polarization (i.e., along the $z$-axis).
The magnitude, however, is about 3 times smaller than that of the x-oriented dipole.

\begin{figure}[b]
\includegraphics[width=\linewidth]{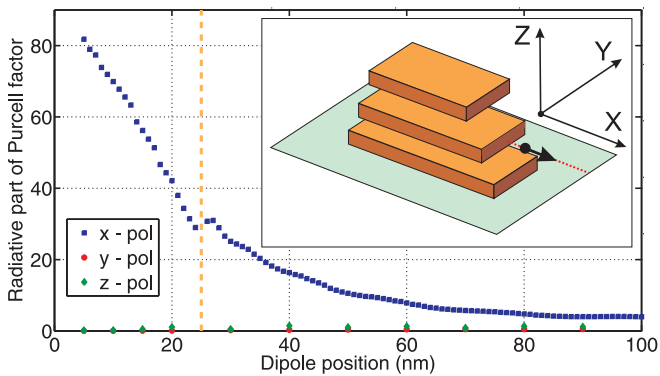}
\caption{(Color online) Calculated dependencies of the radiative part of the Purcell factor
$F_\mathrm{P}^\mathrm{rad}$
on the horizontal distance between the dipole and the edge of the feed element for a $x$-, $y$-, and
$z$-polarized dipole (squares, circles and diamonds, respectively). The vertical dashed line marks
the position exactly above the edge of the reflector. The geometry is explained in the
inset: the $x$-polarized dipole is shown as a black dot with arrow, it is
is centered with respect to the feed element along $y$- and $z$-direction and shifted in $x$-direction, along
the dotted horizontal line.}
\label{Pf_rad}
\end{figure}

In this point let us discuss the results of the full Purcell factor $F_\mathrm{P}$ and the radiative efficiency $\eta$.
The calculated results are shown in Fig.~\ref{Pf}, where the dipole is placed 10~nm above the feed
(i.e., in a homogeneous transparent glass). The Purcell factor is rather large and reaches the maximum of
$F_\mathrm{P} \approx 200 - 300$ when the dipole is located above the edge of the feed element.
Our numerical analysis shows that the growing oscillations of the calculated Purcell factor
when the $x-$coordinate approaches the edges of the director, feed, and reflector elements (vertical dashed lines in Fig.~\ref{Pf})
are the artifacts of our calculation method.
The radiative efficiency appears to be almost independent of the dipole's $x$-coordinate and is about 20\% when the emitter
is above the feed element.

It should be mentioned that we omitted the calculation of the full Purcell factor as well as the radiative
efficiency $\eta$ in Fig.~\ref{Pf_rad}, where the dipole is placed inside a modulated layer. In this case,
it appears that we cannot calculate these quantities correctly in the FMM, because the discontinuous
permittivity function of such a layer is described by a truncated Fourier expansion. Thus, due to the
Gibbs phenomenon, also the non-absorptive surrounding of the metal exhibits a small imaginary part.
The point dipole approximation, however,
fails if the dipole is placed inside a lossy material.

Finally, we would like to mention that the antenna array allows to control the emission directivity
further, by using several emitting dipoles in different positions and controlling their phase difference of oscillations.
Figure~\ref{Dir_Pat}  shows the scaled by $1/2$ 2D polar diagrams of the emission of two synchronized dipoles, positioned at neighboring
antennas of the array at distance $\vec{d}$, proportional to integer numbers of periods
(see dotted and dashed  lines in the left and central panels).
The emission intensity of the synchronized pair of dipoles normalized to the maximum emission intensity of two non-synchronized dipoles
is then simply
 $P_2(\vec{d},\vartheta, \varphi) = P(\vartheta, \varphi)\left[1+\cos \left( 2 \pi n  (\vec{d}\cdot\vec{\mathrm{e}_r})/\lambda \right)\right]$, where $P(\vartheta, \varphi)$
 is the directional pattern of  a single dipole (solid line), and $n$ is the refractive index of superstrate.
 Note that the polar diagrams
 $P_2$ of the synchronized dipoles demonstrate  the effect of super-radiance: the Purcell factor of two synchronized closely positioned
 dipoles becomes twice larger than unsynchronized.
 It is seen also that it becomes possible to increase the directivity using two emitting dipoles.


To conclude, we show that the gold Yagi-Uda nanoantenna array enhances
and simultaneously directs radiation of the oscillating point dipole. The
dipole-nanoantenna enhancement depends very strongly on the oscillating
frequency, dipole position, and orientation of its moment. It becomes possible to
control the emission directivity further by using several synchronized emitting dipoles attached
to different antennas in the array.
This  opens a way to manipulate an excited-state lifetime of a quantum emitter
and to fabricate narrow beaming nanoscale antennas in the optical range.

We acknowledge support from BMBF (13N 9155, 13N 10146), DFG (FOR 557, FOR 730, GI
269/11-1), DFH/UFA, ANR Chair of Excellence Program, the Russian Academy of Sciences,
and the Russian Foundation for Basic Research

\end{document}